# Transport catastrophe analysis as an alternative to a monofractal description: theory and application to financial time series

Sergey A. Kamenshchikov*

*Moscow State University of M.V.Lomonosov, Faculty of Physics,
Russian Federation, Moscow, Leninskie Gory, Moscow, 119991*

*IFC Markets Corp., Analytics division, UK, London,
145-157 St John Street, EC1V 4PY*

*Corresponding e-mail: kamphys@gmail.com*

___________________________________________________________________

**Abstract**

The goal of this investigation was to overcome limitations of a persistency analysis, introduced by Benoit Mandelbrot for monofractal Brownian processes: nondifferentiability, Brownian nature of process and a linear memory measure. We have extended a sense of a Hurst factor by consideration of a phase diffusion power law. It was shown that pre-catastrophic stabilization as an indicator of bifurcation leads to a new minimum of momentary phase diffusion, while bifurcation causes an increase of the momentary transport. An efficiency of a diffusive analysis has been experimentally compared to the Reynolds stability model application. An extended Reynolds parameter has been introduces as an indicator of phase transition. A combination of diffusive and Reynolds analysis has been applied for a description of a time series of Dow Jones Industrial weekly prices for a world financial crisis of 2007-2009. Diffusive and Reynolds parameters shown an extreme values in October 2008 when a mortgage crisis was fixed. A combined R/D description allowed distinguishing of market evolution short-memory and long memory shifts. It was stated that a systematic large scale failure of a financial system has begun in October 2008 and started fading in February 2009.

___________________________________________________________________

**1 Introduction**

In 1955 the American researcher Hassler Whitney has created a mathematical foundation of a modern catastrophe theory – the theory of mapping singularities [1]. It includes investigations of peculiarity classes, that appear for mapping of one two dimensional surface to another one. Whitney has found out two stable types of mappings – types that have not been destroyed after negligible deformations of surfaces or their projections. These types of mappings have been generalized for arbitrary manifolds with dimension up to 10 by Whitney's followers, for example [2]. One of them led to the discrete change of a system's characteristic state – "cusp" catastrophe. It is represented for a one dimensional case at the Fig.1. The multiplicity or uncertainty is maximal in the unstable area of C – vicinity. According to [1] the disruption $D \to E$ appears as a fusion of stable and unstable regimes, marked by ovals. In terms of a bifurcation theory this one-dimensional evolution corresponds to the saddle-node fusion in a phase space. Another type of destabilization is a self-oscillating destabilization, suggested by H. Poincare (1879) in his dissertation thesis. It has been proved by A. Andronov and E. Leontovich in 1939.

*The author declares that there is no conflict of interests regarding the publication of this article.

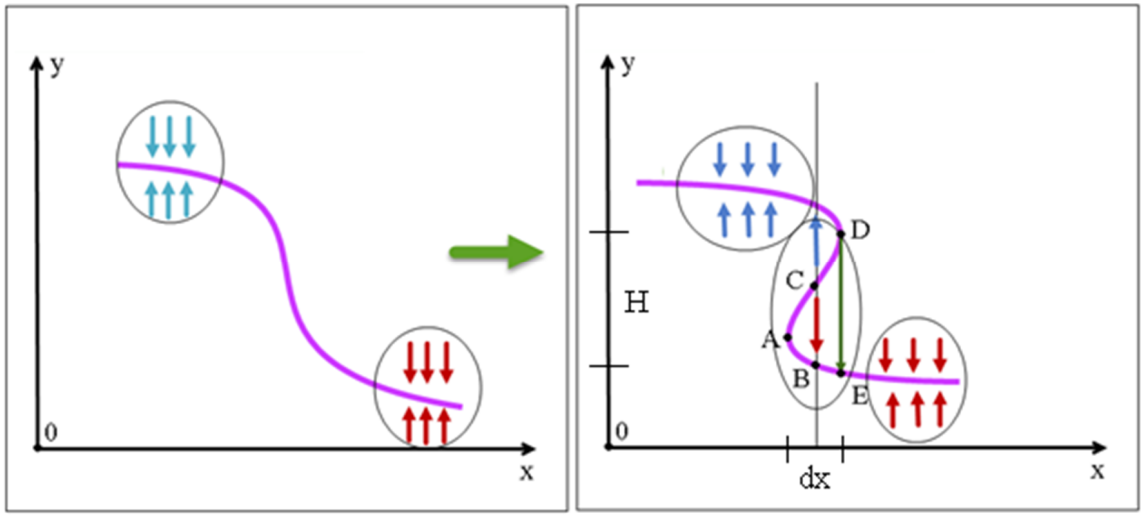

Fig.1. Appearance of Whitney's "cusp"

According to the Andronov-Leontovich theorem [3] a birth of a new limited cycle in a phase space is realized by a transition through a stable equilibrium zone, i.e. the system should return to the stabilization before a new bifurcation occurs. A birth of a new cycle is preceded by the distortion and a death of a previous quasi stable regime.

A "calm before storm" or effect of small scale oscillations suppression before bifurcation has been noticed by M.M. Dubovikov and N.V. Starchenko in [4] as well. They have studied a behavior of financial time series by the use of Hurst monofractal parameter of stability. Anatoly Neishtadt has shown that a delay of the dynamic bifurcation exists in case of all known analytical nonlinear systems [5] for adiabatic change of control parameter. It means that inertial properties resist a new synchronization – the system needs time for the restructuring as it happens in case of Ising model of magnetic domains. A delay depends on the clusters interaction and an intensity of the external "field", i.e. macro scale influence. A comfortable choice of a macroscopic control parameter has been suggested in [6] on basis of Reynolds parameter for turbulent streams:

$$R(t) = f\left(\vec{\Pi}(t)\right) = \frac{q^+(t)}{q^-(t)} \tag{1}$$

Here $R(t)$ is basic phase parameter and $\vec{\Pi}(t)$ is set of microscopic control parameters - parametric vector. Quantities $q^+$ and $q^-$ correspond to power input and output per system unit. In given description bifurcation corresponds to the transition of an equilibrium state $R = 1$:

$$R(t_0) = 1 \Rightarrow \uparrow q^+(t) \Rightarrow \uparrow R(t) \Rightarrow R(t) \succ 1 \Rightarrow \uparrow q^-(t) \Rightarrow \uparrow R(t) \Rightarrow R(t_1) = 1 \tag{2}$$

$$R(t_0) = 1 \Rightarrow \downarrow q^+(t) \Rightarrow \downarrow R(t) \Rightarrow R(t) \prec 1 \Rightarrow \downarrow q^-(t) \Rightarrow \uparrow R(t) \Rightarrow R(t_1) = 1 \tag{3}$$

Here $\uparrow$ and $\downarrow$ show a finite increase and decrease of corresponding parameter for $t_1 \succ t \succ t_0$. The delay between a new cycle appearance and macro scale excitation is defined by the inertial properties of system domains. A new bifurcation has to pass through an equilibrium quasi stable state of $R = 1$.

## 2 Monofractal analysis as an indicator of stability and its limitations

Inability to define strictly a set of control parameters and a global excitation balance obliged researchers to look for statistical measures of the system stability. One of classical approaches is a monofractal analysis, suggested by Benoit Mandelbrot.

According to Mandelbrot and his revolutionary work [7] scale invariance is the necessary property of fractals.

However we should note that chaotic natural systems have scale characteristic limits. For example a turbulent flow has an internal micro scale, defined by inertial viscous forces and an external macro scale, defined by external hydrodynamic influence. Such type of system was denoted as quasi fractal by Mandelbrot, because they have a satisfactory fractal description only within given scale limits $\varepsilon_0 \prec \varepsilon \prec E$. An application of a fractal description to the time series $f(t_i)$ meant an investigation of statistical properties in case of several time resolutions $\varepsilon = \Delta t$.

A single characteristic in frame of a classical monofractal description, introduced by Mandelbrot is a Hurst factor. It was induced through the relation of Fractional Brownian Motion (FBM) [7]. An idea of FBM introduction was inability to explain deviations from normal distributions in some natural systems, for example financial markets. Pareto – Levy distributions have been obtained as particular cases of such abnormal behavior.

Let us consider a Standard Brownian Motion (SBM) time series $B(t)$, which satisfies a normal distribution. Then an FBM increment can be expressed in the following way:

$$B_H(t_2) - B_H(t_1) = \frac{1}{\Gamma(H+1/2)} \int_{-\infty}^{t_2} (s-t_2)^{1/2-H} dB(s) - \frac{1}{\Gamma(H+1/2)} \int_{-\infty}^{t_1} (s-t_1)^{1/2-H} dB(s) \quad (4)$$

A given increment is expressed through the fractal derivatives of SBM with a factor $k = (0.5 - H), 0 \prec H \prec 1$. This factor defines a deviation from a standard markovian Brownian motion ($H = 0.5$). FBH allows obtaining anomalous distributions with "thick tails", and a flexible explaining of flights. According to [8] an expectation of FBM deviation is self-affine:

$$E\left([B_H(t+T) - B_H(t)]^2\right) = V_H \cdot T^{2H} \quad (5a)$$

$$V_H = \left(\frac{1}{\Gamma(H+1/2)}\right)^2 \cdot \left[\int_{-\infty}^{0} \left[(1-s)^{H-1/2} - (-s)^{H-1/2}\right]^2 ds + \frac{1}{2H}\right] \quad (5b)$$

This means that probable amplitude of the deviation depends on a time scale and a Hurst factor $H$ of system's memory. If $H = 0.5$ then relation (5a) corresponds to the Einstein's law of Brownian walks:

$$E\left([B(t+T) - B(t)]^2\right) = V_{0.5} \cdot T \quad (6)$$

If $H \neq 0.5$ we achieve an anomalous transport, that includes Levy flights and "thick tails" of distribution for $H \succ 0.5$. Despite a charm of this approach it has several limitations, enumerated below.

a) <u>FBM is achieved as weighted averaged Brownian motion</u>.
According to the original work of B. Mandelbrot [7] "FBM of the exponent $H$ is a moving average of $dB(t)$ in which past increments of $B(t)$ are weighted by the kernel $(t-s)^{H-1/2}$". The weights are defined on the basis of time distance between current moment and previous states $\Delta = (s - t_j)$. The intensity of a history influence is determined by a memory factor $0 \prec H \prec 1$. However FBM operator assumes SBM kernel for weighted average. It means that FBM is considered as dynamical moving weighted integration of standard Brownian process.

If we calibrate FBM such that $B_H(t_0) = 0$ then an absolute value can be expressed in the following way:

$$_\Delta B_H(t) = \frac{1}{\Gamma(H+1/2)} \int_{-\infty}^{t} (s-t)^{1/2-H} \cdot dB(s) \quad (7)$$

If we consider a motion only in negligible time range $(t-ds,t)$ then this relation can be simplified:

$$dB_H(t) = \frac{1}{\Gamma(H+1/2)} ds^{1/2-H} \cdot B'(s) \cdot ds = \frac{ds^{1/2-H}}{\Gamma(H+1/2)} \cdot dB \qquad (8)$$

Up to a constant factor this relation corresponds to the SBM increment. In such a way FBM assumes limitations of Markovian process that should be satisfied for small time deviations.

b) <u>An increment of the FBM has an infinite exact energy</u>.
As it was stated by Mandelbrot [7] that a first fractal derivative of FBM and consequently its energy diverges for the range $0 \prec H \prec 1$. To overcome this obstacle he has introduced a smoothed derivative where a range of smoothing $\delta$ is defined artificially:

$$B'_H(t,\delta) = -\int_{-\infty}^{\infty} B_H(s) d\varphi(t-s) \qquad (9)$$

$$\varphi(t) = \frac{1}{\delta}, t \in [0,\delta] \quad \varphi(t) = 0, t \prec 0 \quad \varphi(t) = 0, t \succ \delta$$

However that is not the only procedure to introduce "physical" derivative (we may use a weighted derivative as well) and that's why the universality of a dynamic description is lost;

c) <u>Hurst factor expresses a linear measure of memory and is not applicable for nonlinear cases</u>.
This remark needs a certain clarification. According to [7] a linear autocorrelation function of a first derivative can be expressed in the following way:

$$C_H(\tau,\delta) = E\left(\frac{\partial B_H}{\partial t}(t,\delta)\bigg|\frac{\partial B_H}{\partial t}(t+\tau,\delta)\right) = V_H \cdot H \cdot (2H-1)|\tau|^{2(H-1/2)} \qquad (10)$$

It is a quadratic function of Hurst factor and again depends on the artificial smoothing parameter $\delta$. Cases of $H \succ 0.5$ and $H \prec 0.5$ correspond to the persistent and antipersistent trends correspondingly [4,8]. In case of markovian SBM $H = 0.5$ and $C_H(\tau,\delta) = 0$. For markovian process a probability connection is stated by Chapman - Kolmogorov equation which is not linear in general:

$$W(x_3,t_3|x_1,t_1) = \int dx_2 W(x_3,t_3|x_2,t_2) \cdot W(x_2,t_2|x_1,t_1) \qquad (11)$$

In fact a Hurst factor as a memory measure can be applied for the characterization of linear trends in regard to the function $_\Delta B_H(t)$, but according to a standard FBM model it can't be used generally for the indication of pre-catastrophic stabilization, considered in the Section 1.

d) Hurst factor is a single macroscopic fractal characteristic. It means that a detailed structure of a non uniform phase area should be investigated with use of additional parameters. If self affine properties depend on space-time scale then other models can be used for a more subtle description – for example the multifractal detrended fluctuation analysis [9].

Limitations, mentioned above, can be partially overcome by the introduction of extended Hurst analysis.

**3 Extended interpretation of Hurst parameter**

In this section we shall consider a diffusive approach to the characterization of pre-catastrophic stabilization effect (PS-effect), noted in the Section 1. This description assumes an introduction of a second transport factor, used in a standard Fokker-Planck equation (12).

This equation has been derived on a basis of Chapman - Kolmogorov equation (11) and thus is applied for markovian processes:

$$\frac{\partial P(x,t)}{\partial t} = \frac{1}{2} \cdot \frac{\partial}{\partial x}\left(D(x) \cdot \frac{\partial P(x,t)}{\partial x}\right) \quad D(x) = \lim_{\Delta t \to 0}\left(\frac{\langle\langle_\Delta x^2\rangle\rangle}{\Delta t}\right) = \frac{\langle\langle_\Delta x^2\rangle\rangle}{\Delta t_{min}} \quad (12)$$

Here double brackets designate an averaging of an initial coordinate:

$$b(x,_\Delta t) = \int (x-x_0)^2 \cdot W(x,x_0,_\Delta t)dx_0 = \langle\langle_\Delta x^2\rangle\rangle \quad (13)$$

It has been shown in [9] that systems of phase mixing have a second transport factor $D$ with explicit time dependence, expressed through the specific energy of characteristic vector:

$$D(x,t) = \frac{\langle\langle_\Delta x^2\rangle\rangle}{\Delta t_{min}} =_\Delta t_{min} \cdot \int \varepsilon(x,t) \cdot W(x,x_0,_\Delta t)dx_0 =_\Delta t_{min} \cdot \langle\langle\varepsilon(x,t)\rangle\rangle \quad (14)$$

$$\varepsilon(x,t) = \lim_{\Delta t \to 0}\left(\frac{_\Delta x(x,t)}{_\Delta t}\right)^2 = \left(\frac{_\Delta x(x,t)}{_\Delta t_{min}}\right)^2 \quad (15)$$

Let's introduce a variable time lag $T = t - t_0$ and a power regression of the transport factor that will be denoted as a dynamic diffusion below:

$$D(t_0,T) = D_0(t_0) \cdot T^\kappa \quad (16)$$

Then an expectation of the stochastic shift can be represented in the following way:

$$E\left([x(t_0+T) - x(t_0)]^2\right) = D_0(t_0) \cdot T^{\kappa+1} \quad (17)$$

A comparison of this relation with an equation (5a) allows expressing the Hurst factor through a stability coefficient $\kappa$: $H = \frac{\kappa+1}{2}$. Unlike a FBM procedure we made no assumptions regarding a micro scale probability distribution function. That's why a generalized Hurst parameter has no obligatory preliminary limitations in frame of this model $-\infty \prec H \prec \infty$. It doesn't create an artificial energy divergence as well and may be applied to natural systems directly without smoothing. However an extended Hurst analysis still has a limitation of markovian processes (13). A critical value of $H_{cr} = 0.5$ corresponds to the constant diffusion case $\kappa = 0$ and is a boundary of the diffusive expansion $H \succ 0.5$ and the diffusive contraction $H \prec 0.5$ of a characteristic area. Thus a generalized Hurst factor is a measure of the attractor stability. We may introduce a potential of attraction on the basis of a diffusive scale:

$$\Lambda_D = \sqrt{\langle\langle_\Delta x^2\rangle\rangle} = \sqrt{D_0(t_0)} \cdot T^H \quad (18)$$

Then a diffusive acceleration can be represented in the following way:

$$\frac{\partial^2 \Lambda}{\partial T^2} = \sqrt{D_0(t_0)} H(H-1) \cdot T^{H-2} = -\frac{\partial V_{eff}}{\partial x} \quad (19)$$

Here $V_{eff}$ is an efficient volume potential of attraction of a given phase space. In regard to the time series analysis it characterizes a volatility of a considered time series:

$$\int \sqrt{D_0(t_0)} H(1-H) \cdot T^{H-2} dx = V_{eff} \quad (20)$$

It is an integral characteristic of a manifold internal interaction. According to the relation (20) a system may loose memory even for "persistent" case of $H \succ 0.5$.

A sense of a generalized Hurst factor can be clarified with a use of spectral description. If we introduce a characteristic frequency $_\Delta t = 1/f$, then according to the relation (18) a following formula may be represented for a dynamic diffusion spectrum:

$$D(t_0, f) = D_0(t_0) \cdot f^{-\kappa} \quad (21)$$

We may note that the case of $H \succ H_{cr}$ corresponds to the large scale transport, while if $H \prec H_{cr}$ a micro scale energy absorption is more intense. A shift of a basic absorption band from micro scale range to macro scale range corresponds to catastrophic behavior, when a coherent motion and resonances appear.

It should be remarked that in general case of a non uniform phase area the $\kappa$ factor may in its turn depend on the frequency. An additional nonlinear form of the diffusive law then may be introduced. However the relation (21) can still be used as the initial approach for the description of macroscopic diffusive properties.

## 4 Diffusive analysis and R – bifurcations

In Section 1 we have considered a pre-catastrophic stabilization effect (*PS* - effect) as a first necessary condition of the bifurcation. It should be marked that a PS-effect leads to the small scale spectrum band intensification with a following transition to a large scale transport. One of possible ways, that can be used to compare macro/micro transport properties, is represented by the relation (22).

$$I(t_0) = \frac{\int_{T_{\min}}^{T_{\max}/2} D(t_0, T) dT}{\int_{T_{\min}}^{T_{\max}/2} D(t_0, T) dT} \qquad (22)$$

Here an integral stabilization factor *I* is expressed through the relation of small frequency and high frequency integrals. A total integration range $(T_{\min}, T_{\max})$ is defined with an account of measurement resolution.

Another alternative is a momentary transport analysis of the uniform markovian time series $x_i = x(t_i)$. It can be outlined on a basis of the expression (14). This factor (23) allows defining an average transport for the period $T = t_i - t_{i-N}$ and reaches a new minimum during a pre-catastrophic stabilization phase: $_\Delta D(x_i, T) = 0$. A disruption leads to an increase of the momentary transport due to a large scale motion.

$$D(x_i, T) = \frac{\langle\langle _\Delta x_i^2 \rangle\rangle}{t_i - t_{i-N}} = \frac{\sum_{j=i-N}^{i}(x_i - x_j)^2}{T} \qquad _\Delta D(x_i, T) = 0 \Rightarrow _\Delta D(x_i, T) \succ 0 \qquad (23)$$

An alternative choice of a control parameter has been suggested in a Section 1 on the basis of an extended Reynolds factor. It may be denoted as a basic phase parameter $R(t)$. According to mechanisms (2) and (3) a bifurcation corresponds to the state $R(t) \neq 1$, and a following formation of a new quasi cycle such that $R(t) \to 1$. As it was shown in [10] two principle types of disruption are possible. Let's use a dynamic entropy of Kolmogorov $h_d$ [11]: $h_d = \langle h(\vec{x}(t)) \rangle$.

Here averaging in phase space is designated as $\langle \ \rangle$ and averaged quantity can be expressed as sum of positive Lyapunov factors $h_i^+$ for each dimension of generalized phase space:

$$h = \sum_{i=N}^{K} h_i^+ = \ln\left(\prod_{i=N}^{K} \sigma_i^+\right) \qquad \sigma_i^+(t) = \frac{|\delta x_i(t)|}{|\delta x_i^0|} \qquad (24)$$

We may denote $\vec{x}(t)$ is characteristic phase vector of system state. Factor $\sigma_i^+$ shows distance growth $\delta x_i(t)$ in *i* direction for two infinitely closely located points in phase space. Condition of stationary state then is equal to $h = 0$ or $\sigma_i^+ = 1$ ($i = \overline{N, K}$).

Let's introduce a relation for specific system power:

$$q(t) = \frac{\delta}{\delta t}\left(\sum_{i=1}^{K} \frac{v_i^2}{2}\right) = \sum_{i=1}^{K} v_i \cdot \dot{v}_i = q^+(t) - q^-(t) \qquad (25)$$

System stability condition leads to $h_i \leq 0$ and $\sigma_i \leq 1$. Given inequalities lead to expression (26) for velocity components $v_i\ (i = \overline{1,K})$.

$$\frac{|\delta x_i(t)|}{|\delta x_i^0|} = \frac{|\delta x_i(t)/\delta t|}{|\delta x_i^0/\delta t|} = \frac{|v_i(t)|}{|v_i^0|} = \alpha(t) \qquad 0 \prec \alpha(t) \leq 1 \qquad (26)$$

Relation (26) in fact allows receiving components of acceleration $\dot{v}_i(t)$.

$$\lim_{\Delta t \to 0}\left(\frac{|v_i(t)| - |v_i^0|}{\Delta t}\right) \leq 0 \qquad \dot{v}_i(t) = \lim_{\Delta t \to 0}\left(\frac{v_i(t) - v_i^0}{\Delta t}\right) \qquad (27)$$

In such a way a consideration of specific power $q(t)$ can be reduced to two cases: a) $\delta x_i(t) \succ 0$ and $|\delta x_i(t)| = \delta x_i(t)$; b) $\delta x_i(t) \leq 0$ and $|\delta x_i(t)| = -\delta x_i(t)$. Signs of $\delta x_i(t)$ and $\delta x_i^0$ match - this condition is obligatory for definition of Lyapunov factors. Then $\alpha(t)$ doesn't depend on initial sign of coordinate shift $\delta x_i^0$.

For cases a) and b) we then receive: a) $\dot{v}_i \prec 0$ and $v_i \succ 0$; b) $\dot{v}_i \geq 0$ and $v_i \leq 0$. In both cases with use of relation (26) we receive that $q(t) \leq 0$. According to the definition of a basic phase parameter this means that $R(t) \leq 1$ for $h_d \leq 0$. Condition of $R_f$, i.e. $R(t) = 1$ corresponds to a new stable regime. Thus use of $R(t)$ as control parameter must be delimited for two types of system [6]: a) accelerator - $\dot{v}_i(t) \succ 0$; b) decelerator - $\dot{v}_i(t) \leq 0$. For first type of system motion stability loss and bifurcation are realized for $R(t) \prec 1$, while decelerator comes to transition only for $R(t) \succ 1$. In both cases a transition corresponds to the following requirement:

$$R(t) = 1 \Rightarrow R(t) \neq 1 \qquad {}_\Delta\varepsilon(t) = 0 \Rightarrow {}_\Delta\varepsilon(t) \neq 0 \qquad (28)$$

**5 Reynolds and diffusive analysis: application to financial markets**

In this section we shall consider several illustrations of diffusive and *R*-analysis applications to the financial time series. Let's consider a time period of 16.07.2006-21.02.2010 which includes two significant phases – a beginning of the world financial crisis and a gradual recovering. According to the analysis of George Soros [12] a preliminary origin of the crisis corresponds to the falling of bank liquidity in August 2007. In September 2008 it caused a failure of greatest American mortgage agencies: Lehman Brothers, Fannie Mae, Freddie Mac.
Only in January 2009 US Federal Reserve has started the fourth supporting program of financial stabilization (QE4) that led to a preliminary recovering of a financial system.
Let's analyze a time series of DJI weekly prices – an index price was defined at the trading end of each Friday – the end of a trading week. In Fig.2 it is shown that an upward trend, marked as rising corridor, has been broken in August 2007.
A total collapse of this index corresponds to the failure of mortgage agencies in October 2008. Let's demonstrate an application of two approaches that were considered above – analysis of a basic phase parameter $R(t)$, *R* - analysis and a diffusive analysis, *D* - analysis.

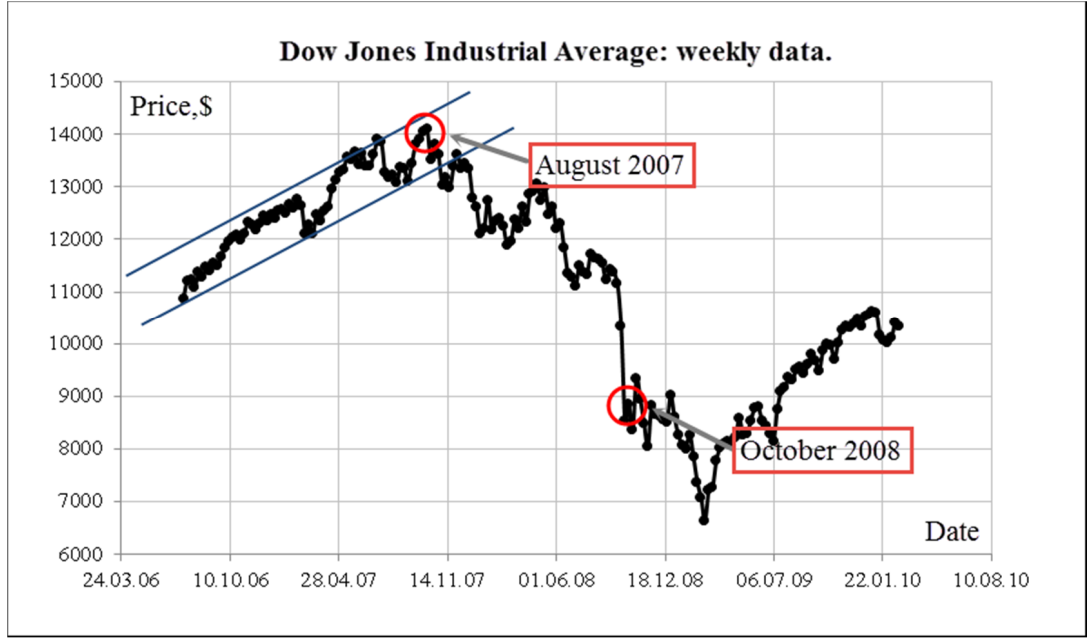

**Fig. 2**. DJI weekly time series. Two stages of crisis – a failure of bank liquidity in August 2007 and the collapse of key mortgage agencies in September 2008.

Both will be applied for the discovering of new system disruptions, corresponding to macro scale shifts of price.

$$\varepsilon_i = \frac{1}{2}\left(\frac{x_i - x_{i-1}}{_\Delta t_i}\right)^2 \quad _\Delta\varepsilon_i = \varepsilon_i - \varepsilon_{i-1} \quad _\Delta t_i =_\Delta t = const \quad i = \overline{1, N} \quad (29)$$

The set of relations (29) expresses a difference approximation of the parameter $_\Delta\varepsilon_i(t_i)$. According to the conditions, considered above, an attraction sign should be defined simultaneously: $\dot{v}_i(t) \succ 0$ or $\dot{v}_i(t) \prec 0$ scenario must be chosen for a correct indication of the bifurcation: a) $\dot{v}_i(t) \succ 0$ and $_\Delta\varepsilon_i(t_i) \prec 0$ or b) $\dot{v}_i(t) \prec 0$ and $_\Delta\varepsilon_i(t_i) \succ 0$. Let's compose a time dependence of a product $Bif_i =_\Delta \varepsilon_i \cdot \dot{v}_i$, where acceleration is expressed by formulas (30).

$$v_i = \frac{x_i - x_{i-1}}{_\Delta t_i} \quad \dot{v}_i = \frac{v_i - v_{i-1}}{_\Delta t_i} \quad i = \overline{1, N} \quad (30)$$

An appearance of a disruption then corresponds to the transition $Bif_i = 0 \Rightarrow Bif_i \prec 0$. A chart of a normalized bifurcation indicator is represented in Fig.3. We can note that a minimal value corresponds to the largest index fall of September-October 2008: the failure of greatest American mortgage agencies and a mortgage crisis in US. Let's mark the date points of a critical bifurcation parameter $Bif_i \prec 0$. They are denoted by red points of date axis in Fig.4. We have marked four clusters of critical points, corresponding to bifurcations: 14.10.07-24.02.2008, 04.05.2008-22.06.2008, 17.08.2008 – 05.10.2008 and 11.01.2000 – 15.02.2009. First cluster defines a delay between a bank crisis and a market reaction – we may observe inertial properties of a market system. However $R$ – analysis shows preliminary signals before a trend channel breakthrough. A third cluster corresponds to the mortgage crisis and a minimal $Bif_i$ of Fig.3.

A second approach that should be considered is a diffusive markovian analysis, represented above. The transport factor approximation has been calculated on the basis of relations (23). Its normalized values are displayed in Fig.4.

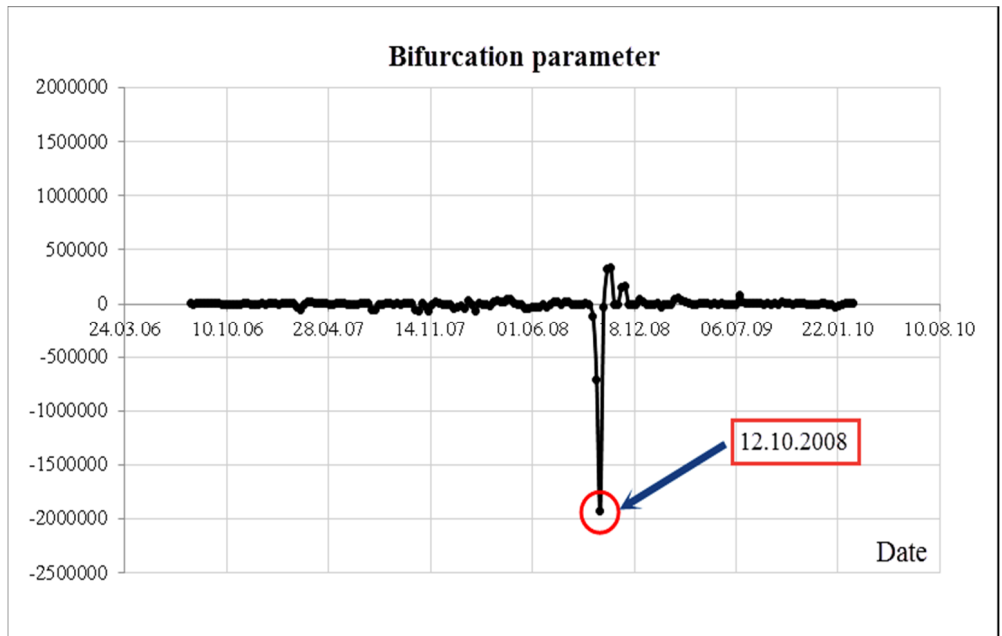

**Fig. 3**. Bifurcation parameter time series: 16.07.2006-21.02.2010.

Highest amplitude of $D_i$ fluctuation again corresponds to October 2008, the mortgage crisis. An each intersection of a date axis indicates a new markovian bifurcation. Points of intersections have been defined through the following condition:

$$_\Delta D(x_{i-1},T) \cdot _\Delta D(x_i,T) \prec 0 \qquad (31)$$

This relation allows determining of derivative sign change: $_\Delta D(x_{i-1},T) \succ 0 \Rightarrow _\Delta D(x_i,T) \prec 0$
$_\Delta D(x_{i-1},T) \prec 0 \Rightarrow _\Delta D(x_i,T) \succ 0$.

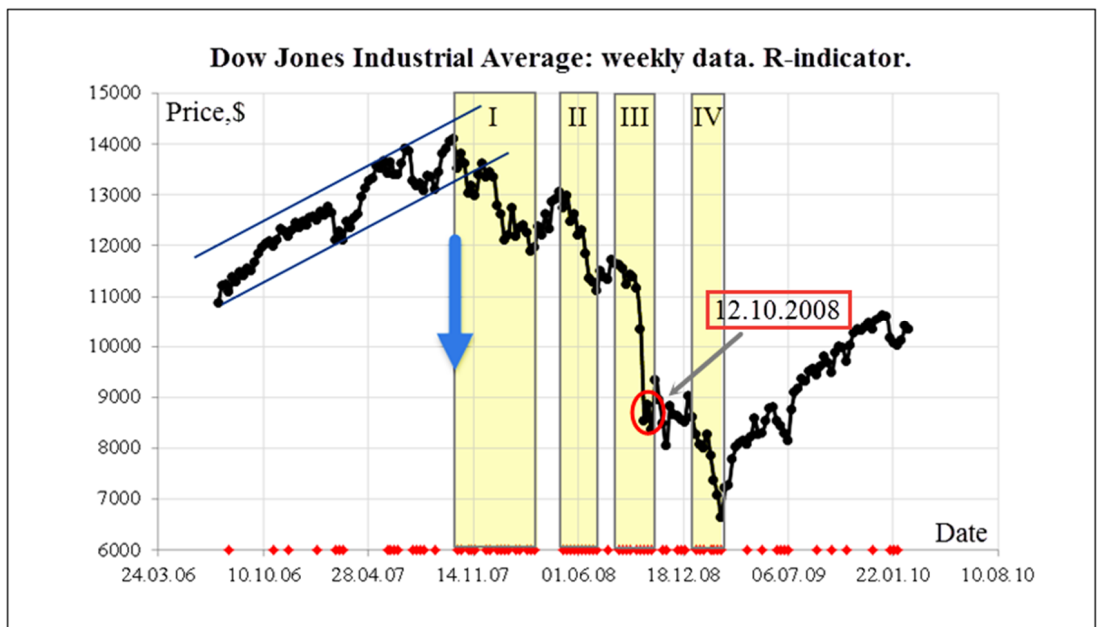

**Fig. 4**. Clusters of a critical bifurcation: $R$ – analysis.

In Fig 6 principal clusters are marked in the OX axis as it has been done in Fig.4. It is significant to emphasize that an $R$ – analysis has a wider area of application since it is not limited by markovian processes and may be applied to processes with long memory, like inertial trends.

In Fig.6 areas of disruption, defined by both *R* – analysis and *D* – analysis, are marked by blue rectangles. Thus a diffusive analysis has demonstrated 50% efficiency in relation to the *R* – description.

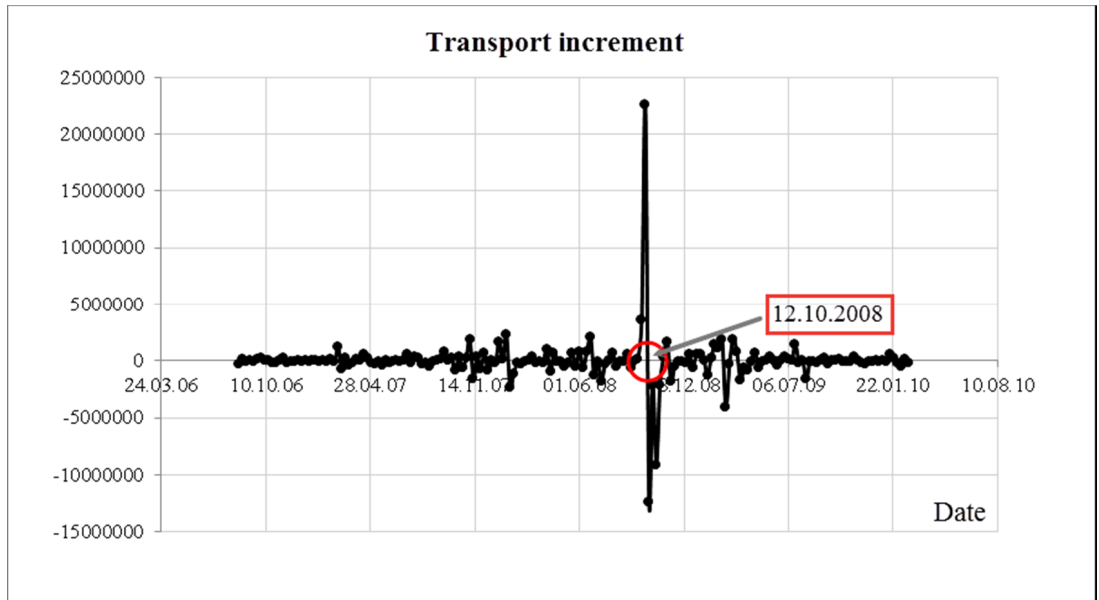

**Fig. 5**. A transport increment time series: 16.07.2006-21.02.2010.

Let's designate a combination of *R* – analysis and *D* – analysis as *R/D* - analysis. This combined type of a description allows distinguishing of short-memory, markovian stages of a market evolution and long memory processes. Considered historical period of 2006-2010 allowed finding out of two short memory periods – 14.10.07-24.02.2008, 04.05.2008-22.06.2008 and two long memory stages – 17.08.2008 – 05.10.2008, 11.01.2009 – 15.02.2009.

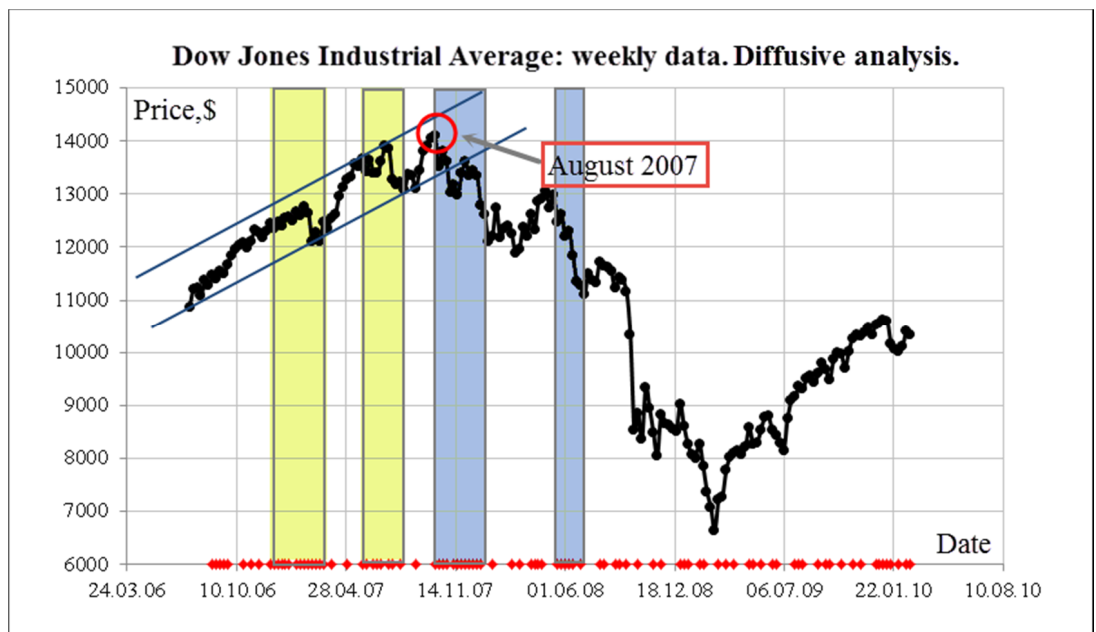

**Fig. 6**. Clusters of a critical bifurcation: diffusive analysis.

If we define a financial crisis as a systematic, inertial large scale failure of basic economic indexes, then, according to *R/D* – analysis, its beginning corresponds to October 2008. This is a date of propagation beginning through adjacent developed markets like markets of European Union.

# 6 Conclusions

In this paper a combined transport analysis has been considered for the description of a pre-catastrophic stabilization. PS - effect has been stated as a fusion of stable and unstable parametric areas and phase cycles. A delay between a new cycle appearance and a macro scale excitation is defined by inertial properties of system domains – an indication of transport anomalies may help to forecast a catastrophe. We have shown that a classical monofractal analysis introduce artificial properties into physical process: nondifferentiability, Brownian nature and linear memory measure. In this frame a classical Hurst factor can not be used for the indication of nonlinear pre-catastrophic stabilization.

A sense of a Hurst factor has been extended by the consideration of a phase diffusion law. This model characterizes a shift of a basic absorption band from a micro scale to a macro scale range when a coherent motion and resonances appear. To indicate this effect momentary phase diffusion has been introduced. It reaches a new minimum during a pre-catastrophic stabilization phase: $_\Delta D(x_i, T) = 0$. A disruption leads to a following increase of the momentary transport due to a large scale motion.

An efficiency of a diffusive analysis has been experimentally compared to the Reynolds stability model application. It has been shown that an $R$ – analysis has a wider area of application since it is not limited by markovian processes and may be applied to processes with long memory, like inertial trends. A combined diffusive and Reynolds analysis has been applied for a description of a time series of Dow Jones Industrial weekly prices during a world financial crisis of 2007-2009. Diffusive and Reynolds parameters shown extreme values in October 2008 - the failure of greatest American mortgage agencies. This combined type of a description has allowed distinguishing of short-memory markovian stages and long memory processes of a market evolution. The considered historical period of 2006-2010 allowed finding out of two short memory periods. It was stated that a systematic large scale failure of a financial market began in October 2008 and started fading in February 2009.